\documentclass[aip,superscriptaddress,preprint]{revtex4-1} 

\usepackage{graphicx}
\usepackage{dcolumn}
\usepackage{cleveref}
\usepackage{bm}
\usepackage{amsmath}
\usepackage[export]{adjustbox}
\usepackage{subcaption}

\begin{document}


\title{Electrically detected magnetic resonance of carbon dangling bonds at the Si-face 4H-SiC/SiO$_2$ interface}


\author{G. Gruber}
\email[]{gernot.gruber@alumni.tugraz.at}
\affiliation{Institute of Solid State Physics, Graz University of Technology, Petersgasse 16, 8010 Graz, Austria}

\author{J. Cottom}
\affiliation{University College London - Department of Physics \& Astronomy, Gower Street, London, WC1E 6BT, United Kingdom}
\author{R. Meszaros}
\affiliation{Infineon Technologies, Siemensstra{\ss}e 2, 9500 Villach, Austria}
\author{M. Koch}
\affiliation{Institute of Experimental Physics, Graz University of Technology, Petersgasse 16, 8010 Graz, Austria}
\author{G. Pobegen}
\affiliation{KAI GmbH, Europastra{\ss}e 8, 9500 Villach, Austria}
\author{T. Aichinger}
\affiliation{Infineon Technologies, Siemensstra{\ss}e 2, 9500 Villach, Austria}
\author{D. Peters}
\affiliation{Infineon Technologies, Schottkystra{\ss}e 10, 91058 Erlangen, Germany}
\author{P. Hadley}
\affiliation{Institute of Solid State Physics, Graz University of Technology, Petersgasse 16, 8010 Graz, Austria}

\date{\today}

\begin{abstract}
SiC based metal-oxide-semiconductor field-effect transistors (MOSFETs) have gained a significant importance in power electronics applications. However, electrically active defects at the SiC/SiO$_2$ interface degrade the ideal behavior of the devices. The relevant microscopic defects can be identified by electron paramagnetic resonance (EPR) or electrically detected magnetic resonance (EDMR). This helps to decide which changes to the fabrication process will likely lead to further increases of device performance and reliability. EDMR measurements have shown very similar dominant hyperfine (HF) spectra in differently processed MOSFETs although some discrepancies were observed in the measured $g$-factors. Here, the HF spectra measured of different SiC MOSFETs are compared and it is argued that the same dominant defect is present in all devices. A comparison of the data with simulated spectra of the C dangling bond (P$_\textrm{bC}$) center and the silicon vacancy (V$_\textrm{Si}$) demonstrates that the P$_\textrm{bC}$ center is a more suitable candidate to explain the observed HF spectra.
\end{abstract}


\maketitle


\section{Introduction}

Silicon carbide (SiC) is a wide band gap semiconductor with material properties suitable for high power, high temperature, and high frequency applications. However, while much research and development of SiC devices has been carried out in the past decades, there is still room for improving the performance of the devices. In the case of SiC metal-oxide-semiconductor field-effect transistors (MOSFETs), the channel mobility remains more than one order below the bulk value.\cite{bib:RozenTED} Furthermore, complex threshold voltage variations are present in modern SiC MOSFETs.\cite{bib:RescherMSF2016, bib:RescherIEDM} It is well established that there is a high density of interface traps ($D_\textrm{it}$) present at the SiC/SiO$_2$ interface of the most common polytypes \cite{bib:AfanasevPSSA162} and that passivation can be achieved by nitridation, particularly by post oxidation anneals (POAs) in a nitric oxide (NO) atmosphere.\cite{bib:RozenTED,bib:SalinaroTED62} However, there is no clear consensus on the microscopic structure of the dominant electrically active defects. While the dominant interface defects in Si MOSFETs have been identified with electric paramagnetic resonance (EPR) decades ago,\cite{bib:NishiJJAP,bib:LenahanJVST16} for SiC this is not the case. Numerous studies that have attempted to identify those defects in SiC devices by means of EPR and electrically detected magnetic resonance (EDMR) are summarized in the work by T. Umeda \it et al. \rm\cite{bib:UmedaECS} In recent work performed on the 4H-SiC/SiO$_2$ interface, two candidate defects have frequently been discussed. The first one is the carbon dangling bond (P$_\textrm{bC}$) center \cite{bib:CantinPRL92, bib:MeyerAPL86, bib:IsoyaMSF2002, bib:MacfarlaneJAP88, bib:CantinAPL88, bib:BardelebenMSF2004, bib:UmedaMSF2011} and the second one is the silicon vacancy (V$_\textrm{Si}$).\cite{bib:MeyerAPL84, bib:DautrichAPL89, bib:CochraneJAP109, bib:CochraneAPL100, bib:CochraneAPL102, bib:AichingerAPL101, bib:AndersTED62} In this study the EDMR spectra obtained from differently processed 4H-SiC $n$-channel MOSFETs are compared and simulated spectra based on the reported hyperfine (HF) parameters of the P$_\textrm{bC}$ and V$_\textrm{Si}$ defects are discussed. While the dominant defect in the studied devices has been tentatively assigned to the V$_\textrm{Si}$ in previous studies,\cite{bib:GruberMSF2016,bib:GruberICDS} the comparison to the simulations demonstrates that the P$_\textrm{bC}$ center is a more suitable candidate for the observed interface defect.

\subsection{What can be learned from EPR/EDMR?}

The EDMR method is a well established technique to identify paramagnetic defect centers in semiconductors and has successfully been used for the identification of defects in fully processed SiC devices.\cite{bib:CochraneAPL100,bib:CottomJAP} EDMR is related to EPR and takes advantage of the fact that a portion of the current through a semiconductor device may be spin dependent.\cite{bib:StutzmannJNCS} In this work, spin dependent recombination (SDR) was measured. Recombination of carriers is most efficient through defect levels deep in the band gap.\cite{bib:ShockleyPR87,bib:HallPR87} If such a defect state is paramagnetic and an external magnetic field $B$ is applied the recombination rate through the defect is decreased.\cite{bib:KSM} By applying a suitable microwave field, the paramagnetic defect can be brought to resonance resulting in an increase of the recombination rate which can be observed as a change in the recombination current. This allows for the measurement of the EPR spectrum of the defects in a device by monitoring the current.\cite{bib:StutzmannJNCS}

\begin{figure}[b]
\includegraphics[width=\columnwidth,max width=10cm]{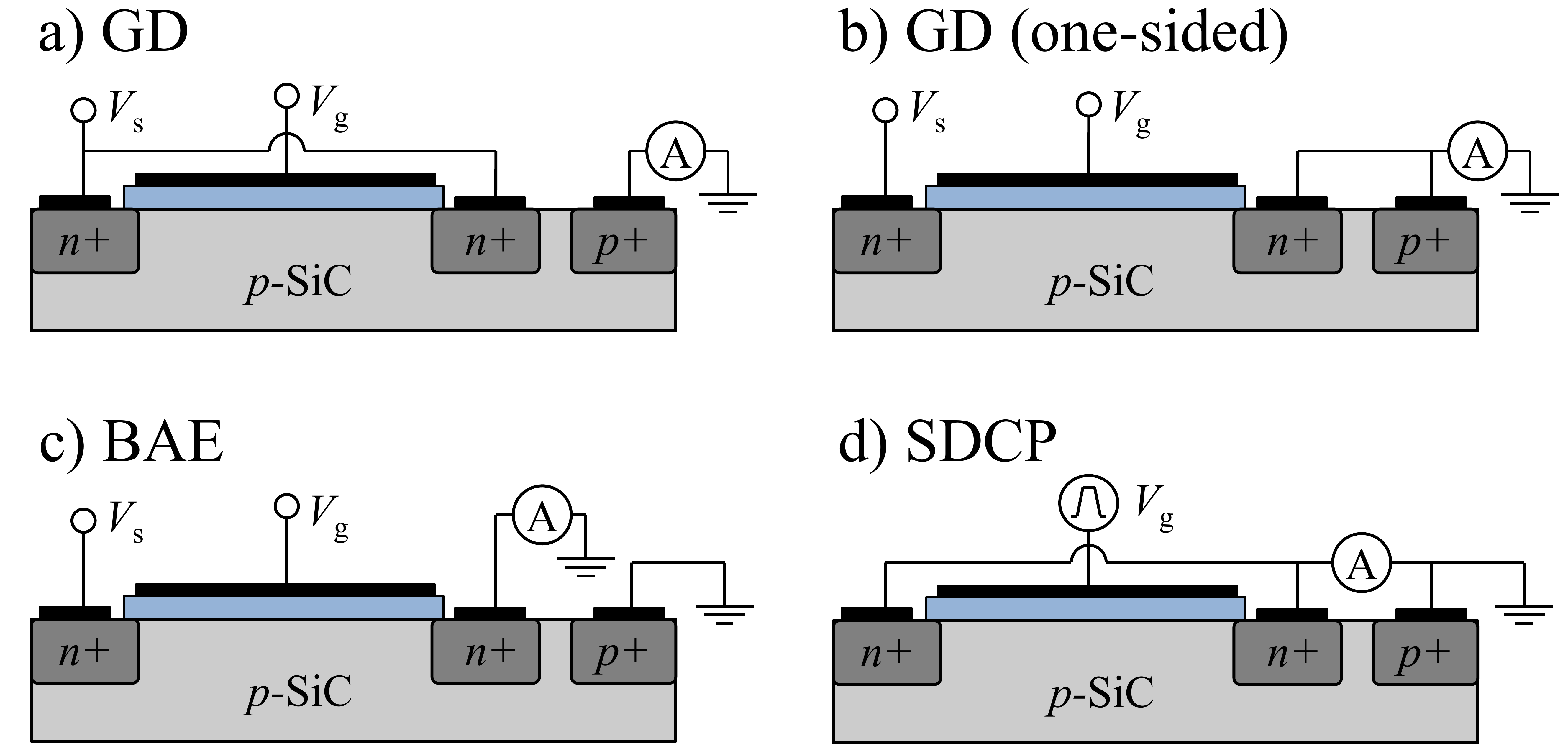}
\caption{Schematic of different biasing schemes for SDR measurements on fully functional MOSFETs, here shown for the example of an $n$-channel MOSFET. Note that the samples are also exposed to a suitable magnetic field and microwave radiation during the measurement. (a) Basic GD biasing scheme,\cite{bib:Jupina, bib:AichingerAPL101} (b) adapted GD for MOSFETs where the source and body are internally shorted,\cite{bib:GruberICDS} (c) BAE technique,\cite{bib:AichingerAPL101} and (d) SDCP technique.\cite{bib:BittelAPL}}
\label{fig:all_SDR}
\end{figure}

Different biasing schemes that can be used for the detection of the EDMR spectrum of interface defects in fully functional MOSFETs are depicted in Fig. \ref{fig:all_SDR}(a)-(d). In all these biasing schemes, the electrons from the $n^+$-region(s) and holes from the $p$-body region are brought to recombination at the interface region. The recombination is monitored by a current measurement. In the gated diode (GD) technique,\cite{bib:Jupina, bib:AichingerAPL101} as shown in Fig. \ref{fig:all_SDR}(a), the source voltage $V_\textrm{s}$ is used to inject electrons from the $n^+$-regions into the $p$-body. The gate voltage $V_\textrm{g}$ is used to establish a situation where the injected electrons have a high probability to recombine with holes through deep level defects at the semiconductor-oxide interface, which is usually in depletion when $n\approx p$. The GD method has successfully been used for the identification of interface defects,\cite{bib:Jupina} but has some drawbacks. The signal-to-noise ratio is diminished by the relatively high bulk current masking the small current change when the studied interface recombination centers become resonant. In addition, bulk defects residing in the space charge region of the $pn$-junction may add to the observed spectrum. For MOSFETs with internally shorted source and body contacts, the GD biasing has to be adapted to one-sided biasing,\cite{bib:GruberICDS} as shown in Fig. \ref{fig:all_SDR}(b). While in principle the measurement and drawbacks remain the same, the carriers are now only injected through one $n^+$-region. Note that there is an additional gate-dependent offset current, due to the short-circuit between drain and body.\cite{bib:GruberICDS} A method with a much increased sensitivity with respect to GD is the bipolar amplification effect (BAE) method,\cite{bib:AichingerAPL101} as shown in Fig. \ref{fig:all_SDR}(c). In this technique electrons are injected from the source $n^+$-region and detected as a current at the drain $n^+$-region, while the body current is ignored. This results in a dramatic increase in the signal-to-noise ratio and selectivity to interface defects, as the parasitic effects described for GD are avoided.\cite{bib:AichingerAPL101} An alternative technique which is also very sensitive to recombination centers at the interface is spin-dependent charge pumping (SDCP),\cite{bib:BittelAPL} as shown in Fig. \ref{fig:all_SDR}(d). While the MOSFET is operated like a gated-diode, $V_\textrm{g}$ is pulsed between full inversion and accumulation, which alternately fills the interface region with electrons and holes. Any carriers that get trapped at interface defects during a semi-pulse may recombine with carriers of the opposite charge when the opposite semi-pulse arrives. The resulting current is highly dependent on the recombination rates of interface defects and can be used for very sensitive EDMR measurements.\cite{bib:BittelAPL}

The spectroscopic information that can be gained from an EDMR spectrum as discussed in this work is contained in the resonance condition, which for the case of one unpaired electron is

\begin{equation}
h\nu=g\mu_\textrm{B}(B+\sum_ka_km_{I,k})
\label{eq:resonance}
\end{equation}

\noindent where $h$ is Planck's constant, $\nu$ is the microwave frequency, $g$ is the $g$-factor, $\mu_\textrm{B}$ is the Bohr magneton, $a_k$ is the HF splitting constant of the $k$-th nucleus and $m_{I,k}$ is the magnetic nuclear spin quantum number of the \mbox{$k$-th} nucleus and $k$ sums over the nuclei interacting with the electron. The $g$-factor is dependent on the spin-orbit coupling and by measuring the angular dependence of $g$ on the direction in which $B$ is applied to the crystal one can study the symmetry of the defect. The HF interaction results in a shift of the resonance magnetic field, which is expressed by the sum in equation \eqref{eq:resonance}. Also this  interaction can have an angular dependence. For the defects considered in this work, the HF structure is caused by the interaction of the unpaired electrons with $^{13}$C and $^{29}$Si atoms with a nuclear spin of $I=1/2$ in both cases. The former have a natural abundance of $1.1\,\%$ and the latter $4.67\,\%$, which is reflected in the relative intensity of the HF lines in the spectrum. Comparing an experimental spectrum to defect models with known HF splittings (from theoretical calculations or from other experiments) is an efficient way to interpret the spectrum of an observed defect. Note that equation \eqref{eq:resonance} only describes the resonance of one individual defect with a given set of $a_k$ and $m_{I,k}$, while an experimental spectrum contains the sum off all possible permutations. The number of individual lines can be very high but there is an efficient method to generate an accurate spectrum from known HF parameters of the nuclei involved, as described in a related study.\cite{bib:CottomJAP} In this approach the total spectrum is generated by a sum of derivative Lorentzians of equal linewidths. Every line has a resonance field resulting from the $a_k$ and $m_{I,k}$ values of the involved nuclei and a relative intensity proportional to the probability to find the set of nuclei in the respective $m_{I,k}$ states. A computer code is used to find the line positions and relative intensities of all lines that have a significant contribution to the spectrum while ignoring the enormous number of lines with very little probability, i.e. lines that contain a high number of the low-abundant spin 1/2 isotopes of Si and C. All remaining lines are then added together to result in the complete simulated spectrum.\cite{bib:CottomJAP}

\subsection{Previous EPR/EDMR measurements at the 4H-SiC/SiO$_2$ interface}

While many different defect models for defects at the SiC/SiO$_2$ interface have been proposed in the literature,\cite{bib:UmedaECS} in recent work two defects have frequently been suggested to be dominant in EPR/EDMR: i) the P$_\textrm{bC}$ center and ii) the negatively charged V$_\textrm{Si}$. 

\begin{figure}[t]
\includegraphics[width=0.65\columnwidth,max width=10cm]{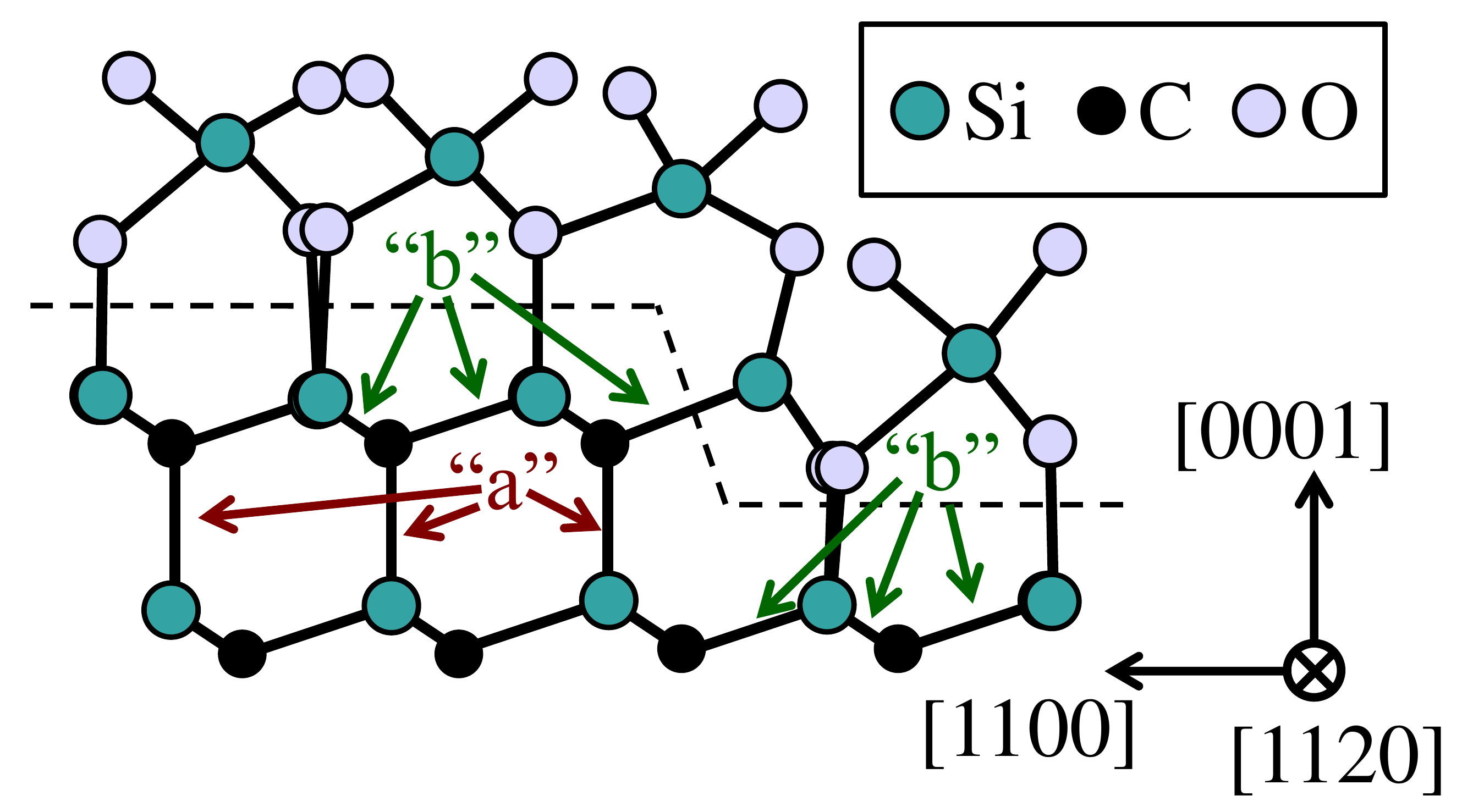}
\vspace{-.3cm}
\caption{A schematic model of the bonding structure at  the Si-face SiC/SiO$_2$ interface (indicate by the dashed line) with axial (``a'') and basal (``b'') C bonds indicated.}
\label{fig:model}
\end{figure}

The P$_\textrm{bC}$ center is well characterized in an EPR study on oxidized porous SiC by J.L. Cantin \it et al.\rm\cite{bib:CantinPRL92} In that study the $g$-factor of the differently oriented dangling bonds at the various interfaces was determined to be $g_\parallel=2.0023$ when the magnetic field $B$ is applied along the symmetry axis of a dangling bond and $g_\perp\approx2.0032$ in the perpendicular direction. The HF parameters for the P$_\textrm{bC}$ center are $a_{\textrm{C},\parallel}\approx80\,\textrm{G}$ and $a_{\textrm{C},\perp}\approx38\,\textrm{G}$ for the central C atom and $a_\textrm{Si}\approx13\,\textrm{G}$ for its neighboring Si atoms.\cite{bib:CantinPRL92} Fig. \ref{fig:model} shows a schematic of the bonding structure at the Si-face SiC/SiO$_2$ interface. C bonds labeled ``axial'' are aligned with the crystalline $c$-axis, while those labeled ``basal'' are not. Note that the axial bonds points towards the bulk SiC and are therefore less likely to be broken on the Si-face, while in an oxidized porous SiC sample all variations are present.\cite{bib:CantinPRL92} 

The negatively charged V$_\textrm{Si}$ defect in bulk SiC is well characterized by an isotropic $g$-factor of $g\approx2.0028$.\cite{bib:WimbauerPRB56,bib:MizuochiPRB68} The HF parameters due to the four neighboring C atoms are $a_{\textrm{C},\parallel}\approx28\,\textrm{G}$ with $B$ applied in the symmetry direction of the unsaturated C bond and $a_{\textrm{C},\perp}\approx10.5\,\textrm{G}$ in the perpendicular direction, as well as $a_\textrm{Si}\approx3\,\textrm{G}$ for the twelve next neighbor Si atoms.\cite{bib:MizuochiPRB68} Several EDMR studies linked the observed spectrum to the V$_\textrm{Si}$ defect, predominantly based on the isotropic $g$-factor.\cite{bib:MeyerAPL84, bib:DautrichAPL89, bib:CochraneJAP109, bib:CochraneAPL100, bib:CochraneAPL102, bib:AichingerAPL101, bib:AndersTED62} Note that the reported values are spread over a range of $g\approx2.0023-2.0031$.\cite{bib:DautrichAPL89,bib:CochraneAPL102} The study by C.J. Cochrane \it et al. \rm resolved the HF structure of the V$_\textrm{Si}$ using a \it fast passage \rm EDMR measurement.\cite{bib:CochraneAPL100} However, the other referenced studies used a conventional detection scheme (without \it fast passage\rm) and showed a somewhat different HF spectrum.\cite{bib:MeyerAPL84, bib:DautrichAPL89, bib:CochraneJAP109, bib:CochraneAPL102, bib:AichingerAPL101, bib:AndersTED62} A recent study by M.A. Anders \it et al. \rm ruled out the presence of dangling bond defects at the SiC/SiO$_2$ interface.\cite{bib:AndersAPL2016} However, their argumentation was based on the absence of resonance lines additional to the dominant defect spectrum which was assigned to the V$_\textrm{Si}$ based on its isotropic $g$-factor. Nonetheless, the difference in the observed HF spectra with and without \it fast passage \rm hints at a different dominant defect in the respective measurement. Only those defects with long spin relaxation times are probed by the \it fast passage \rm EDMR.\cite{bib:CochraneAPL100} Defects not meeting this criterion are not probed with \it fast passage\rm, while they can still be the dominant recombination defect observed in conventional EDMR.

\section{Experimental}

\subsection{Sample description}

The devices discussed in the following have already been part of previous EDMR studies and all have an intentional high EDMR signal and poor device performance.\cite{bib:GruberMSF2016, bib:GruberICDS} All devices are $n$-channel MOSFETs fabricated on the Si-face of 4H-SiC wafers with a $4\,^\circ$ offset with respect to the crystalline $c$-axis and are summarized in table \ref{tbl:devices}. The first device is a MOSFET that received a state-of-the-art oxide deposited by chemical vapor deposition (CVD) and received a 20\,min POA in an O$_2$ atmosphere at $1100\,^\circ\textrm{C}$. This short anneal was necessary to assure a good contact of the oxide on the substrate while it does not passivate interface defects. This MOSFET was specifically designed for the application of the BAE \cite{bib:AichingerAPL101} and SDCP \cite{bib:BittelAPL} methods. The device was compared to identically processed devices with different POA atmospheres in previous studies.\cite{bib:SalinaroTED62, bib:GruberMSF2016} It was concluded that this device contains the same dominant EDMR active interface defect as identically processed devices that received POAs in a N-containing atmospheres.\cite{bib:GruberMSF2016} The second device is a MOSFET with a thermally grown oxide. It received a POA in an N$_2$O atmosphere at $1280\,^\circ$C and was also characterized in a related study.\cite{bib:GruberMSF2016} Also this device allows for the application of the BAE technique. The third device is a MOSFET with the geometry of a double-diffused MOSFET (DMOS) with a thermally grown oxide that received a POA in an N$_2$O atmosphere at $1280\,^\circ$C. Since there was no seperate body contact it only allowed for the application of the less sensitive GD SDR technique, as described in.\cite{bib:GruberICDS}

\begin{table}[t]
\caption{Processing parameters of the studies SiC MOSFETs and observed $g$-factors.\cite{bib:GruberMSF2016,bib:GruberICDS}}
\label{tbl:devices}
\begin{ruledtabular}
\begin{tabular}{lllll}
Sample & Oxide process &  $g_{B\parallel c}$ & $g_{B\perp c}$ \\
\hline
Dep. w/ O$_2$ & CVD + POA (O$_2$, 1100$\,^\circ$C) & 2.0042(4) & 2.0017(4) \\
Therm. w/ N$_2$O & Thermal (N$_2$O, 1280$\,^\circ$C) & 2.0036(4) & 2.0026(4) \\
DMOS w/ N$_2$O & Thermal (N$_2$O, 1280$\,^\circ$C) & 2.0051(4) & 2.0029(4) \\
\end{tabular}
\end{ruledtabular}
\end{table}

\subsection{Comparison of the different samples}

\begin{figure}[b]
\centering
\begin{subfigure}{\columnwidth}
\includegraphics[width=\columnwidth,max width=10cm]{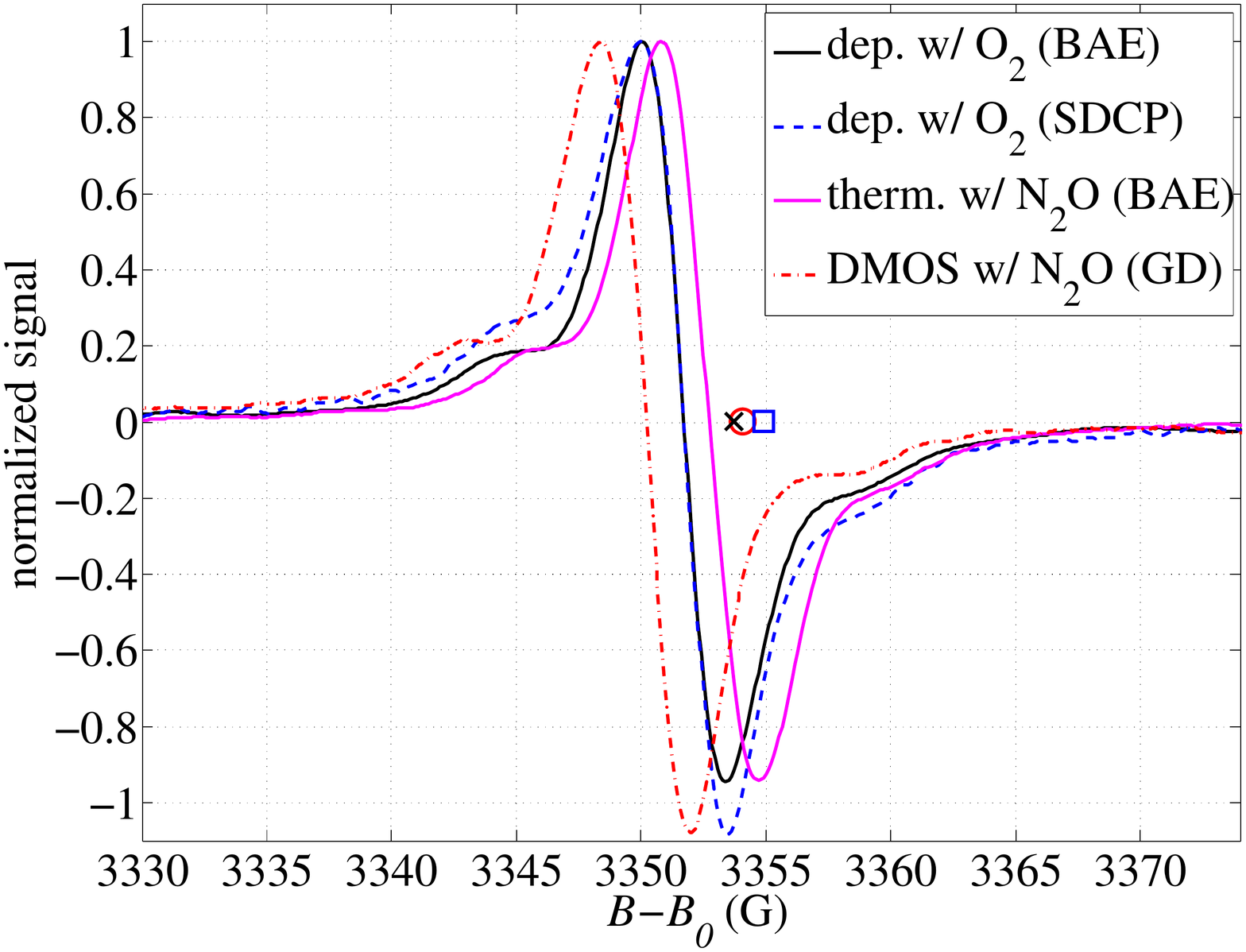}
\vspace{-.8cm}
\caption{Comparison of the spectra as measured.}
\label{fig:all_g}
\end{subfigure}
\begin{subfigure}{\columnwidth}
\includegraphics[width=\columnwidth,max width=10cm]{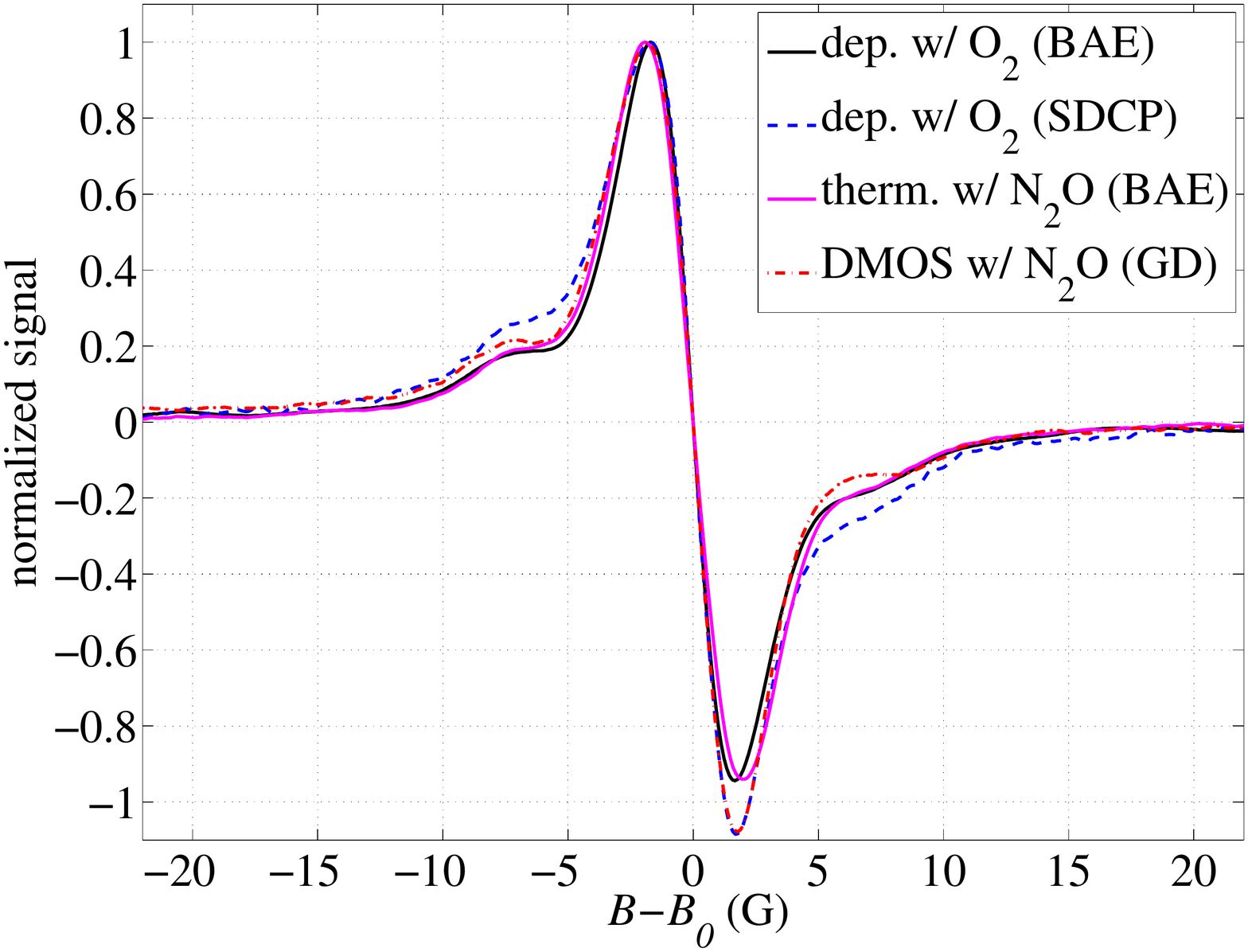}
\vspace{-.8cm}
\caption{Comparison of the HF peaks.}
\label{fig:all}
\end{subfigure}
\caption{Comparison of the normalized experimental EDMR spectra of different devices with $B\parallel c$. The respective EDMR detection technique labeled in parentheses (BAE,\cite{bib:AichingerAPL101} SDCP,\cite{bib:BittelAPL} and GD \cite{bib:GruberICDS}). The respective $g$-factors are listed in table \ref{tbl:devices}. (a) Shows a comparison of the curves as measured compared to the expected positions of the V$_\textrm{Si}$ (red circle), the basal P$_\textrm{bC}$ (black cross), and the axial P$_\textrm{bC}$ (blue square) from the literature.\cite{bib:MizuochiPRB68, bib:CantinPRL92} (b) Shows the curves shifted to center field for a comparison of the HF peaks.}
\label{fig:all_MOS}
\end{figure}

The recorded EDMR spectra with $B$ applied in the crystalline $c$-direction are shown in Fig. \ref{fig:all_MOS}. Note that the device dep. w/ O$_2$ was measured by BAE and SDCP. Fig. \ref{fig:all_g} shows the spectra recorded with a microwave frequency of $f_\textrm{mw}\approx9.402\,\textrm{GHz}$ and compares them to the $g$-factors of the V$_\textrm{Si}$ and P$_\textrm{bC}$ defects. Despite the differences in the observed $g$-factors as listed in table \ref{tbl:devices}, the spectra have a remarkable similarity in the observed HF structure, which can more clearly be seen in Fig. \ref{fig:all}. While not all low intensity HF features are resolved it is evident that all spectra contain a dominant pair of sidepeaks at $\approx\pm6\,\textrm{G}$ from the center line with approximately equal relative intensity. The relative intensity is significantly smaller than that of the identified N$_\textrm{C}$V$_\textrm{Si}$ defect in bulk SiC \cite{bib:CottomJAP} but is very similar, if not identical, to what was observed in comparable EDMR studies of the SiC/SiO$_2$ interface.\cite{bib:AndersAPL2016, bib:MeyerAPL84, bib:DautrichAPL89, bib:CochraneJAP109, bib:CochraneAPL102, bib:AichingerAPL101, bib:AndersTED62} However, those studies reported on an isotropic $g$-factor in the range of $g\approx2.0023-2.0031$.\cite{bib:DautrichAPL89,bib:CochraneAPL102} It is not clear at this point why there is such a discrepancy between the observed $g$-factors. What adds to the problem is that the difference between the expected line positions of the V$_\textrm{Si}$ and P$_\textrm{bC}$ for any orientation is significantly smaller than the observed linewidth. Nonetheless, the defects have significantly different HF parameters. Therefore, we focus on an understanding of the HF structure, as the study of the $g$-factors is inconclusive.

\subsection{Comparison to simulated spectra of the P$_\textrm{bC}$ and V$_\textrm{Si}$}

\begin{figure}[t]
\includegraphics[width=\columnwidth,max width=10cm]{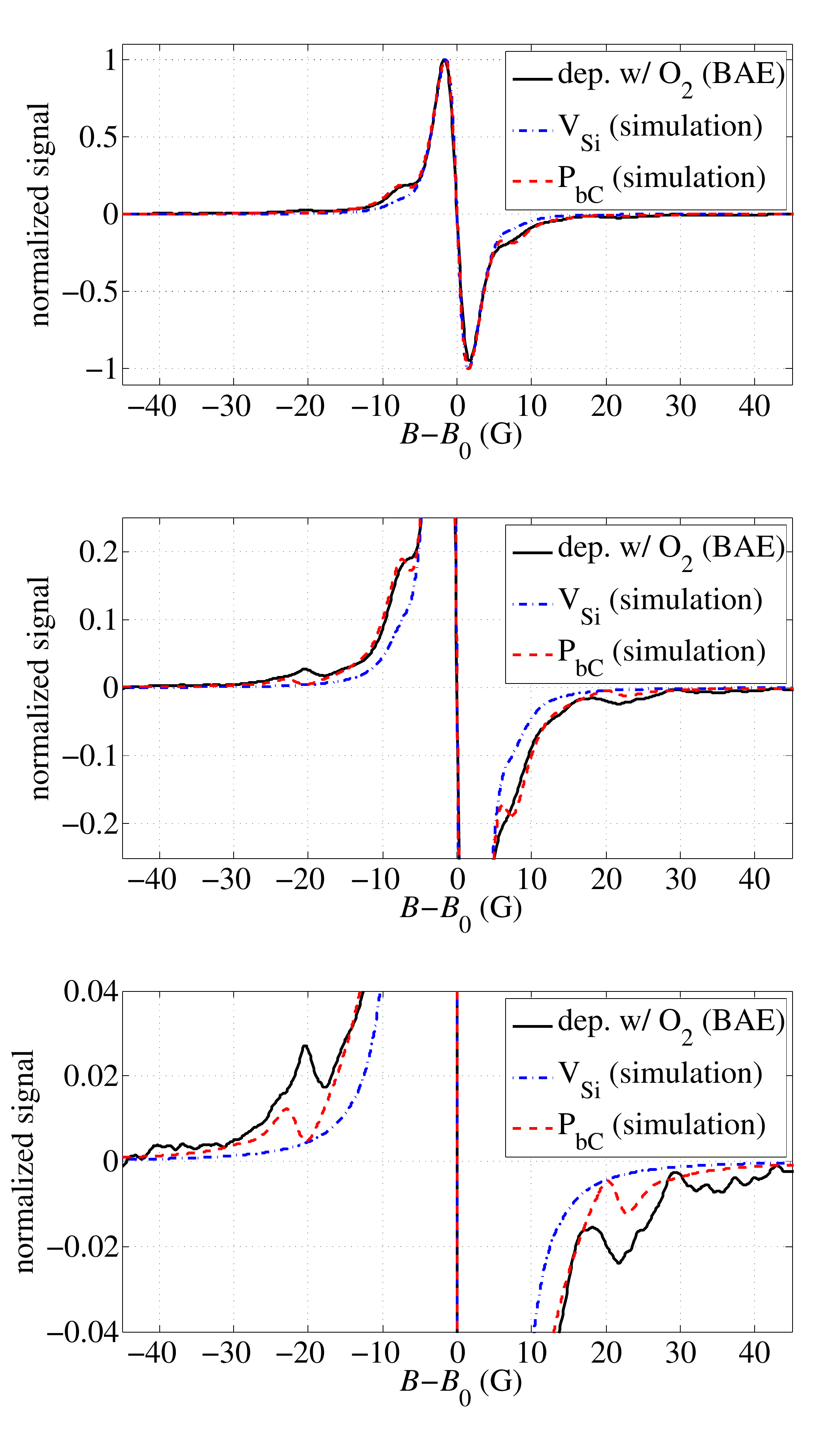}
\vspace{-.8cm}
\caption{Comparison of the experimental EDMR spectrum of sample dep. w/ O$_2$ obtained by BAE with simulated spectra of the $\textrm{V}_\textrm{Si}$ and the basal $\textrm{P}_\textrm{bC}$ defects. The $\textrm{V}_\textrm{Si}$ was simulated with the HF parameters $a_{\textrm{C},1-3}=13\,\textrm{G}$, $a_{\textrm{C},4}=28\,\textrm{G}$ and $a_{\textrm{Si},1-12}=3\,\textrm{G}$ \cite{bib:MizuochiPRB68} and the $\textrm{P}_\textrm{bC}$ with the parameters $a_{\textrm{C}}=43\,\textrm{G}$ and $a_{\textrm{Si},1-3}=13\,\textrm{G}$ .\cite{bib:CantinPRL92} The simulations are composed of a sum of derivative Lorentzians of equal linewidth (matched to the experimental spectrum) with their positions and relative intensities determined by their HF data. The computer code used to generate the spectra is described in a related study.\cite{bib:CottomJAP}}
\label{fig:EDMR_sim}
\end{figure}

For a meaningful comparison of the measured HF structure with simulations it was crucial to resolve as many line features as possible. Out of the measurements described above, the BAE spectrum of sample dep. w/ O$_2$ has the narrowest linewidth and a high signal-to-noise ratio. Since this sample did not receive any passivation by nitrogen it was deemed to contain the highest fraction of defects intrinsic to the SiC-SiO$_2$ system. Additionally, the spectrum was free of small overlapping line features that were observed in some of the nitrited samples.\cite{bib:GruberMSF2016} The spectrum was recorded with measurement parameters chosen to achieve a narrow linewidth and a high signal-to-noise ratio. The measurement was conducted using a microwave frequency of $f_\textrm{mw}\approx9.402\,\textrm{GHz}$ with a nominal power of $P_\textrm{mw}=150\,\textrm{mW}$ as well as a magnetic field modulation at a frequency of $f_\textrm{mod}\approx900\,\textrm{Hz}$ and an amplitude of $B_\textrm{mod}=1\,\textrm{G}$. The signals of 470 individual recordings were averaged resulting in the spectrum which is shown in Fig. \ref{fig:EDMR_sim}. The simulated spectra of the basal P$_\textrm{bC}$ and V$_\textrm{Si}$ were generated by the computer code described in a related study \cite{bib:CottomJAP} using the HF data from the literature.\cite{bib:CantinPRL92,bib:MizuochiPRB68} The spectrum contains two pairs of lines symmetric around the center line. One is at $\approx\pm6\,\textrm{G}$ and the second is at $\approx\pm19\,\textrm{G}$ with a smaller relative intensity. It is evident that the HF features in the experimental spectrum are well represented by the simulation of the P$_\textrm{bC}$ center, despite slightly smaller HF splittings, while the V$_\textrm{Si}$ model does not contain a sufficient relative intensity in its sidepeaks. Using the P$_\textrm{bC}$ model the $\approx\pm19\,\textrm{G}$ lines are explained by the central C atom while the $\approx\pm6\,\textrm{G}$ lines are caused by the three neighboring Si atoms. The good agreement between the simulation and the measurement of the HF spectrum strongly suggests that the observed dominant defect is the P$_\textrm{bC}$ center.

\section{Discussion}

What is not well understood is why the anisotropy of $g$ is different for the different devices. However, all devices shown in the present study possess an anisotropy with $g_{B\parallel c}>g_{B\perp c}$, as listed in table \ref{tbl:devices}. When comparing this anisotropy to the data shown in the study by J.L. Cantin \it et al. \rm \cite{bib:CantinPRL92} it is worth noting that the basal C dangling bonds also have an anisotropy with $g_{B\parallel c}>g_{B\perp c}$. Note that those dangling bonds are also the ones with a HF splitting of $a_\textrm{C}=43\,\textrm{G}$ when $B$ is parallel to $c$, as used for simulation shown in Fig. \ref{fig:EDMR_sim}. The axial C dangling bonds have the opposite anisotropy with $g_{B\parallel c}<g_{B\perp c}$ and a HF splitting of $a_\textrm{C}=80\,\textrm{G}$. No HF pair near $\pm40\,\textrm{G}$ from the center line was observed in this work or in related studies.\cite{bib:GruberICDS,bib:GruberMSF2016} As discussed above, the absence of the axial C dangling bonds is expected due to the bonding structure of the Si-face 4H-SiC/SiO$_2$, which is shown in Fig. \ref{fig:model}. However, while basal P$_\textrm{bC}$ centers can at least qualitatively explain the observed anisotropy, they do not explain its variation between the different samples. Consequently, there is at least one effect that adds to the observed $g$-factor, as discussed below.

i) The first explanation is the presence of an additional defect with a different $g$-factor that adds to the spectrum. If such a defect was present one would expect an influence on the observed $g$-factor dependent on the relative signal of this defect. However, one would also expect a variation of the intensity ratio of the observed HF peaks and a distortion of the central peak. While Fig.  \ref{fig:all} shows some small variations between the samples, there is evidently the same dominant spectrum present in all devices which can be well explained by the P$_\textrm{bC}$ model.

ii) The $g$-factor varies between samples with different oxide growth processes, as the ones shown in this work, while samples using the same oxide growth process but a variation of POAs showed the same $g$-factors.\cite{bib:GruberMSF2016} The older devices that received a thermal oxidation may have a less abrupt or more disordered interface region resulting in less anisotropy as a higher fraction of axial C dangling bonds may be present. However, in addition to the arguments in i) the absence of the $a_\textrm{C}=80\,\textrm{G}$ indicates that predominantly basal C dangling bonds are present.

iii) Variations of the interface abruptness for the different samples may also induce strain to the bonds at the interface. Additionally, there may be strain induced from the variations in the geometry of the stacking structures for the different samples. Strain usually results in a distribution of $g$-factors for one defect in a given direction which induces broadening to the sample, despite shifting the observed zero-crossing.\cite{bib:Weil} While such an effect has been observed in an EDMR study on SiC $p$-channel MOSFETs \cite{bib:AndersTED62} no reliable quantification of this effect was obtainable for the samples studied in the present work. However, due to the large variations on the $g$-factors between the samples one would expect a significant distortion in the observed lineshapes, or at least significant line broadening, which is not observed.

iv) The current used for the EDMR measurement may induce a local magnetic field additional to the applied magnetic field. As the studied devices have different geometries, this may result in differences in the observed $g$-factors. However, for each sample the observed $g$-factor was independent of the current direction or magnitude which is why this effect is excluded.

v) In fully manufactured SiC MOSFETs as studied in this work nickel is used for the ohmic contacts. Ni is ferromagnetic and may perturb the local magnetic field at the defect sites. The differences of the device geometry would result in a variation of this effect, as is observed. Unfortunately, while the influence of the Ni seems to be a very reasonable explanation for the spread in $g$-factors for SiC MOSFETs, a systematic study of this effect using specifically prepared samples was not possible in this work. Also a quantification of this effect is challenging which is why it can only be speculated at this point how much this effect may add to the observations. 

\section{Summary}

In summary, this work demonstrates that the dominant HF spectrum frequently observed in EDMR studies of the Si-face 4H-SiC/SiO$_2$ interface can be understood in terms of P$_\textrm{bC}$ centers. Different devices from different generations of SiC MOSFETs all show very similar HF spectra while they show different magnitudes in the anisotropy of the $g$-factor. While the varieties of the $g$-factor are not well understood, the experimentally observed HF spectrum shows a good match with a simulation of the P$_\textrm{bC}$ center using literature based HF data. The absence of the $a_\textrm{C}=80\,\textrm{G}$ doublet suggests that predominantly basal C dangling bonds are present, which can be explained by the bonding structure on the Si-face of SiC. It was shown before that those interface defects are passivated by anneals in an NO atmosphere while the electrical behavior of the devices significantly improves.\cite{bib:GruberMSF2016} This suggests that a further understanding of interface P$_\textrm{bC}$ centers and their passivation by NO anneals or alternative processes could be valuable for the improvement of device performance and reliability of SiC MOSFETs.

\section{Acknowledgements}

This work was jointly funded by the Austrian Research Promotion Agency (FFG, Project No. 846579) and the Carinthian Economic Promotion Agency Fund (KWF, contract KWF-1521/26876/38867).

\end{document}